\documentclass{moriond}

\def\be{\begin{equation}}
\def\ee{\end{equation}}
\def\bea{\begin{eqnarray}}
\def\eea{\end{eqnarray}}

\usepackage{amsfonts}
\usepackage{amssymb}
\usepackage{amsmath}

\newcommand{\Photo}{}

\begin{document}
\vspace*{4cm}
\title{Inverse Seesaw from dynamical $\mathbf{B-L}$ breaking}

\author{ Julia Gehrlein }

\address{Departamento  de  F\'{\i}sica Te\'{o}rica,  Universidad  Aut\'{o}noma  de  Madrid,\\
Cantoblanco  E-28049  Madrid,  Spain\\
Instituto  de  F\'{\i}sica  Te\'{o}rica  UAM/CSIC,\\
 Calle Nicol\'{a}s Cabrera  13-15,  Cantoblanco  E-28049  Madrid,  Spain}

\maketitle\abstracts{
The Inverse Seesaw scenario relates the smallness of the  neutrino masses to a small $B-L$ breaking parameter.
We investigate a possible dynamical generation of the Inverse Seesaw neutrino mass mechanism from the spontaneous breaking of a gauged $U(1)_{B-L}$.
To obtain an anomaly free theory we need to introduce additional fermions which exhibit an interesting phenomenology. Additionally, we predict a $Z'$ boson associated to the broken $B-L$ which preferentially interacts with the dark sector formed by the extra fermions making it particularly \it{elusive}.
}

\section{Introduction}
The observation of  non-zero neutrino masses is so far the only laboratory-based evidence for physics beyond the Standard Model. Nevertheless, there are many ways to generate neutrino masses. The probably simplest way is via the Seesaw framework \cite{Minkowski:1977sc,Ramond:1979py,GellMann:1980vs,Yanagida:1979as,Mohapatra:1979ia,Schechter:1980gr} where the smallness of the neutrino masses is explained with a large suppression of the electroweak scale.
However, this is not the only    possibility  to   explain
the  smallness  of  neutrino  masses.  The inverse seesaw (ISS) \cite{Mohapatra:1986bd} 
relies on the fact that neutrino masses are protected by the $B-L$ global  symmetry.  If
this  symmetry  is  only  mildly  broken,  neutrino  masses  will  be   suppressed  by
this  small $B-L$-breaking  parameters.  On the other hand,  production  and  detection  of  the
extra ISS  right-handed  neutrinos  at  colliders  as  well  as  their  indirect  effects  in  flavour  and
precision electroweak observables are not protected by this symmetry hence leading to a rich and interesting phenomenology.

To explore a possible dynamical origin of the ISS pattern we choose  to gauge $B-L$ which then gets spontaneously broken \cite{DeRomeri:2017oxa}.

\section{The model}
In the ISS the masses of the light neutrinos are given by
\begin{equation}
  m_\nu\sim v^2 Y_\nu M_N^{-1}\mu (M_N^T)^{-1} Y^T_\nu. 
\end{equation}
With TeV-scale right handed neutrinos and $\mathcal{O}(1)$ Yukawa couplings  $\mu\sim\mathcal{O}$(keV).
Since a hierarchy of mass scales $\mu/v_H\sim 10^{-6}$ seems to be rather ad hoc we will  promote $\mu$ to a dynamical quantity by gauging $B-L$ and identify the $\mu$ parameter with the $B-L$ breaking scalar vev.

For the ISS two copies of right-handed neutrinos ($N_R,~N_R'$) with $B-L$ charges $+1$ and $-1$ per active neutrino are introduced. 


\begin{table}
\caption{Neutral fermions and singlet scalars with their $U(1)_{B-L}$ charge and their multiplicity. 
$\phi_{1,2}$ are SM singlet scalars while $N_R$, $N'_R$ and $\chi_R$ are right-handed and $ \chi_L$ and $\omega$ are left-handed SM singlet fermions respectively.}
\vspace{0.2cm}
\centering
\begin{tabular}{| c| c| c| c| c| c| c| c| c|}
\hline
  Particle & $ \phi_1$ & $ \phi_2$ & $ \nu_L $& $ N_R$ & $ N'_R$ & $ \chi_R$ & $\chi_L$& $\omega$\\ 
  \hline
  $U(1)_{B-L}$ charge & $ +1$ & $+2$ &$-1 $& $-1$ & $+1$ & $+5$ & $+4$& $+4$ \\
  \hline
  Multiplicity & $1$ & $1$ &$ 3 $& $3$ & $3$ & $1$ & $1$ & $1$\\
  \hline
\end{tabular}
\label{tab:particles}
\end{table}

Requiring an anomaly-free theory leads to the introduction of additional chiral fermion content to the model.
We find a phenomenologically interesting and viable scenario for the particle content displayed in tab.~\ref{tab:particles}.
The Lagrangian in in the neutrino sector is then given by
\begin{align}
- \mathcal{L}_\nu &=  \bar L Y_\nu \widetilde{H}  N_R +  {\overline{N_R^c}} M_N N_R' +  \phi_2  \overline{N_R^c} Y_N N_R + \phi_2^* \overline{(N_R')^c}\, Y'_N N_R'  +\phi_1^*  {\overline{\chi_L}} \, Y_\chi \chi_R +  {\rm h.c.}.
\end{align}

The scalar potential of the model can be written as
\begin{align}
  V&=\frac{m_H^2}{2} H^\dagger H+\frac{\lambda_H}{2} (H^\dagger H)^2 + \frac{m_1^2}{2} \phi_1^*\phi_1 + \frac{m_2^2}{2} \phi_2^*\phi_2
      + \frac{\lambda_1}{2}(\phi_1^*\phi_1)^2 + \frac{\lambda_2}{2}(\phi_2^*\phi_2)^2 \\
      &\quad+ \frac{\lambda_{12}}{2}(\phi_1^*\phi_1)(\phi_2^*\phi_2)+\frac{\lambda_{1H}}{2}(\phi_1^*\phi_1)(H^\dagger H)+\frac{\lambda_{2H}}{2}(\phi_2^*\phi_2)(H^\dagger H)-\eta(\phi_1^2\phi_2^*+\phi_1^{*2}\phi_2).\nonumber
\end{align}

Minimalisation of the potential yields a  vev for $\phi_2$
\begin{equation}
  v_2\simeq \frac{\sqrt{2}\eta v_1^2}{m_2^2}~,
\end{equation}
which will be identified with the $\mu$ parameter. To obtain $v_2\sim\mathcal{O}$(keV), one could have $m_2\sim 10~ {\rm TeV}$, $v_1\sim 10~ $TeV, 
and $\eta\sim 10^{-5}$~GeV.

With the conventions $\phi_j=(v_j+\varphi_j+i\, a_j)/\sqrt{2}$ the mixing  angles $\alpha_1$ and $\alpha_2$ between $h-\varphi_1$ and $\varphi_1-\varphi_2$, respectively, are given by 
\begin{equation}
  \tan\alpha_1\simeq \frac{\lambda_{1H}}{\lambda_1}\frac{v}{2v_1},\quad{\rm and}\quad
  \tan\alpha_2\simeq2\frac{v_2}{v_1}.
  \label{eq:tanalpha}
\end{equation}
With $v_1\sim $ TeV and the quartics $\lambda_1$ and $\lambda_{1H}$ are $\mathcal{O}(1)$, the mixing $\alpha_1$ is expected to be small but non-negligible. Due to this mixing the couplings of the Higgs to gauge bosons and fermions get diminished. Relative to the SM couplings they are  
\begin{equation}
  \kappa_F=\kappa_V=\cos\alpha_1,
\end{equation}
which is constrained to be  $\cos{\alpha_1}>0.92$ (or equivalently $\sin{\alpha_1}<0.39$) \cite{Khachatryan:2016vau}.

The $Z'$ gauge boson associated to the broken $B-L$ obtains a mass
 \begin{equation}
 M_{Z'}= g_{\rm BL} \sqrt{v_1^2+4v_2^2}\simeq g_{\rm BL} v_1~.
 \end{equation}
As the largest particles with the largest $B-L$ charges are in the dark sector of the model (the massless Weyl fermion $\omega$ and $\chi_R,~\chi_L$ which will form a Dirac pair), the $Z'$ is very elusive and has a large BR of $70\%$ to invisibles.

The Dirac fermion $\chi$ is stable since it is protected by an accidential $U(1)$ symmetry in the dark sector, hence it is a good DM candidate. Its  main annihilation channels are $\chi \bar{\chi}\to f \bar f$ via the $Z'$ boson exchange 
and $\chi\bar{\chi}\to Z' Z'$ - if kinematically allowed.

The DM $Z'$ interaction leads to a spin-independent scattering in direct detection experiments. With the current 
experimental bound on the spin-independent cross section  we can derive a lower bound on the vev of $\phi_1$:
\begin{equation}
v_1 ~\text{[GeV]}> \left(\frac{2.2\cdot 10^9}{\sigma^{\rm DD}_{\chi}~\text{[pb]}} \right)^{1/4}~.
\end{equation}
This bound pushes the DM mass to be $m_\chi \gtrsim$ TeV. 
For example, for $g_{\rm BL} = 0.25$ and $m_{Z'} = 10$ TeV, a DM mass $m_\chi = 3.8$ TeV is needed to have 
  $\sigma^{\rm DD}_{\chi} ~\sim 9 \times 10^{-10}$ pb. In turn, this bound translates into a lower limit on the vev of $\phi_1$: $v_1 \gtrsim 40$ TeV (with $Y_\chi \gtrsim 0.1$).

 Since the main annihilation channel of $\chi$ is via the $Z'$ which couples dominantly to the dark sector, the bounds from indirect detection searches turn out to be subdominant.

The massless fermion $\omega$  contributes to the relativistic degrees of freedom in the early universe. 
Comparing  the Hubble expansion rate of the Universe with the interaction rate leads to a freeze out temperature of $\omega$ (for values  that satisfy the correct DM relic abundance) of
$T^{\rm f.o.}_\omega \sim 4$~GeV, before the QCD phase transition.

This means that  the SM bath will heat significantly after $\omega$ decouples which suppresses the contribution of $\omega$ to the number of degrees of freedom in radiation:
\begin{align}
\Delta N_{\rm eff} \approx 0.026
\end{align}
which is one order of magnitude smaller than the current uncertainty on $N_{\rm eff}^{exp}=3.04\pm 0.33$ \cite{Ade:2015xua}. Nevertheless, this deviation from $N_{\rm eff}$ could be detected by EUCLID-like survey \cite{Basse:2013zua,Amendola:2016saw} and would be 
an interesting probe of the model in the future. 

More details on the model can be found in \cite{DeRomeri:2017oxa}.

\section{Summary of the results}
\begin{figure}
\centering
\includegraphics[scale=0.5]{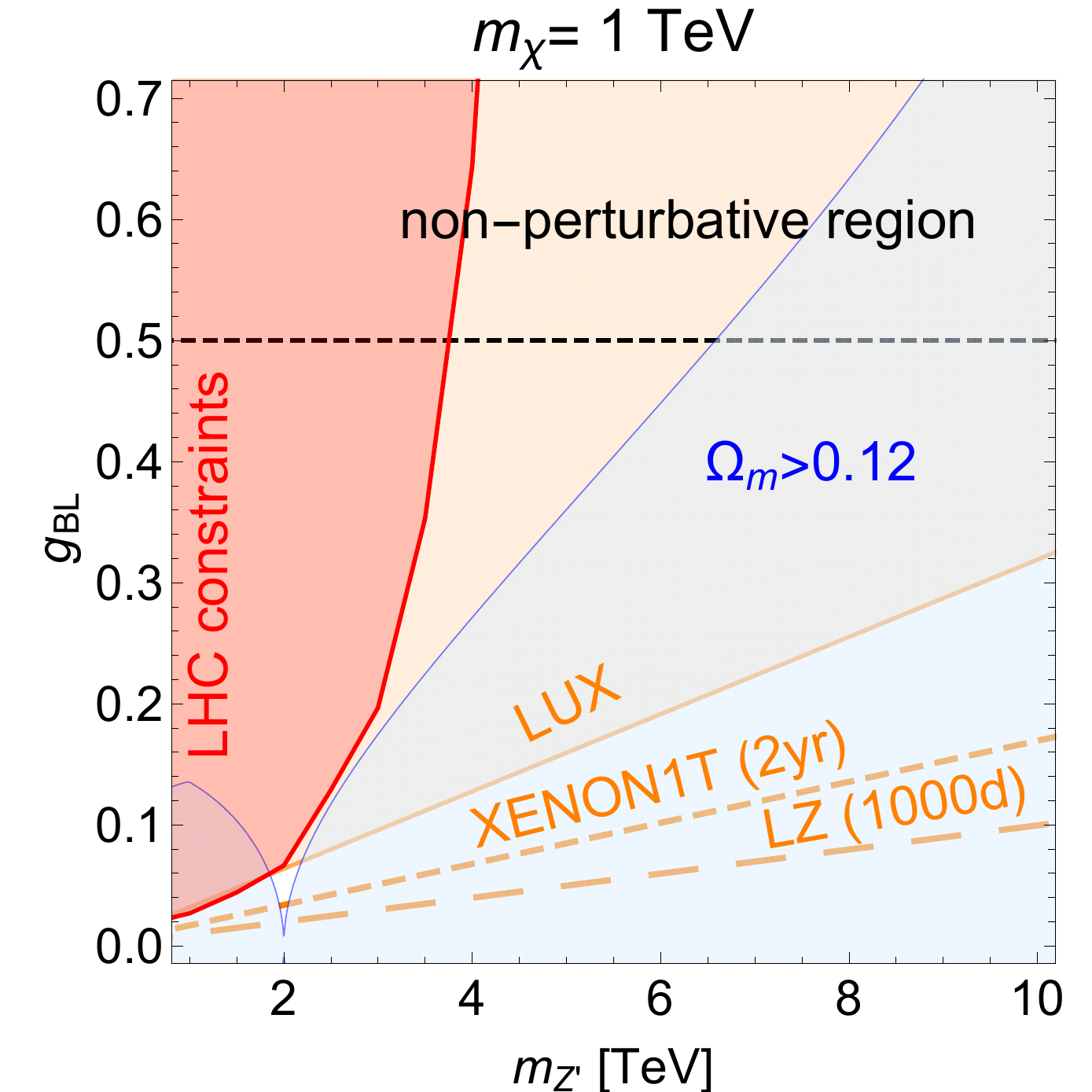}
\includegraphics[scale=0.5]{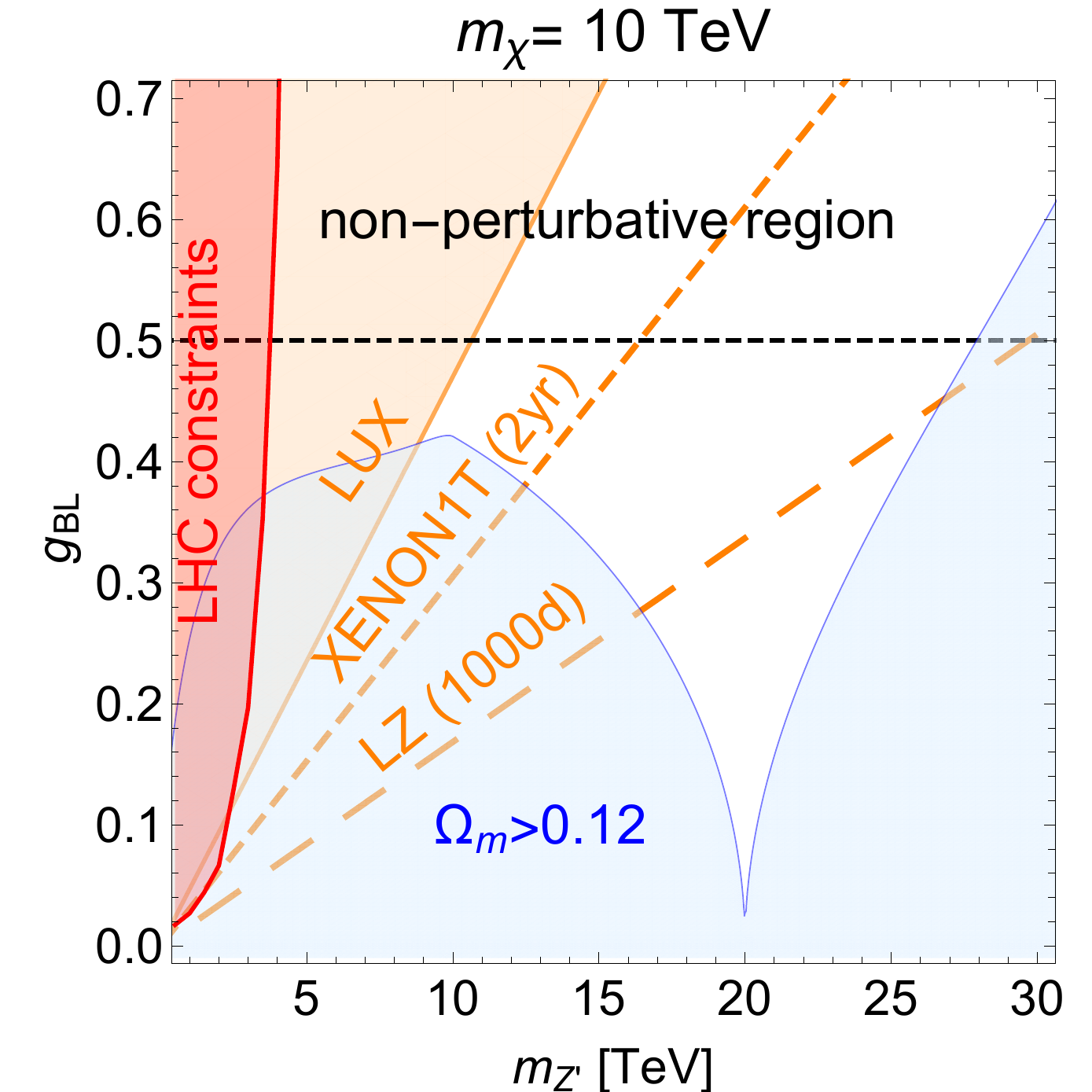}
\caption{\label{fig:g_vs_mz} Summary plots of our results for DM masses of $m_\chi=1 $ TeV and 10 TeV.  The red region to the left is excluded by LHC constraints on the $Z'$, the region above $g_{\rm BL}>0.5$ is 
non-perturbative due to  $g_{\rm BL}\cdot q_{\rm max}\leq\sqrt{2\pi}$. In the blue shaded region DM is overabundant. The orange coloured region is already excluded by direct 
detection constraints from LUX, the short-dashed line indicates the future constraints from 
XENON1T (projected sensitivity assuming $2t \cdot y$), the long-dashed line the future constraints from LZ
(projected sensitivity for 1000d of data taking).}
\end{figure}

Our results  are summarised in fig.~\ref{fig:g_vs_mz}   for different DM masses of $m_\chi=1 $ TeV and 10 TeV.

The red region to the left is constrained by  recasted LHC  $Z'\to e^+e^-,\mu^+\mu^-$ resonant searches~\cite{Khachatryan:2016zqb,ATLAS:2016cyf}. 

In the blue region the relic abundance is too large and the orange region displays the excluded regions from current direct detection experiments \cite{Akerib:2016vxi}, the dashed lines show the sensitivity of near future experiments (XENON1T \cite{Aprile:2015uzo} (projected sensitivity assuming $2t \cdot y$), the long-dashed line the future constraints from LZ \cite{Akerib:2015cja} 
(projected sensitivity for 1000d of data taking)). For gauge couplings above the grey dashed line perturbativity will be lost ($g_{\rm BL}\cdot q_{\rm max}\leq\sqrt{2\pi}$ with largest $B-L$ charge $q_{\rm max}=5$).

In summary, we have presented a dynamical realisation of the inverse seesaw scenario which predicts a 
DM candidate at the TeV scale which can lead to the correct relic abundance  while evading  all current direct and indirect detection constraints, a massless fermion which contributes to $N_{eff}$, an elusive $Z'$ at the TeV scale and two scalars with masses around 10 TeV. Additionally, our model exhibits the phenomenology of the right-handed neutrinos of the usual inverse seesaw.

\section*{Acknowledgments}

 JG is supported by the EU grants H2020-MSCA-ITN-2015/674896-Elusives. JG thanks the organisers of Moriond 2018 EW for the  financial
support that allowed her to attend the conference.

\section*{References}
\bibliography{paper_IFT}

\begin{thebibliography}{10}

\bibitem{Minkowski:1977sc}
Peter Minkowski.
\newblock {$\mu \to e\gamma$ at a Rate of One Out of $10^{9}$ Muon Decays?}
\newblock {\em Phys. Lett.}, B67:421--428, 1977.

\bibitem{Ramond:1979py}
Pierre Ramond.
\newblock {The Family Group in Grand Unified Theories}.
\newblock In {\em {International Symposium on Fundamentals of Quantum Theory}
  {and Quantum Field Theory Palm Coast, Florida, February 25-March 2, 1979}},
  pages 265--280, 1979.

\bibitem{GellMann:1980vs}
Murray Gell-Mann, Pierre Ramond, and Richard Slansky.
\newblock {Complex Spinors and Unified Theories}.
\newblock {\em Conf. Proc.}, C790927:315--321, 1979.

\bibitem{Yanagida:1979as}
Tsutomu Yanagida.
\newblock {HORIZONTAL SYMMETRY AND MASSES OF NEUTRINOS}.
\newblock {\em Conf. Proc.}, C7902131:95--99, 1979.

\bibitem{Mohapatra:1979ia}
Rabindra~N. Mohapatra and Goran Senjanovic.
\newblock {Neutrino Mass and Spontaneous Parity Violation}.
\newblock {\em Phys. Rev. Lett.}, 44:912, 1980.

\bibitem{Schechter:1980gr}
J.~Schechter and J.~W.~F. Valle.
\newblock {Neutrino Masses in SU(2) x U(1) Theories}.
\newblock {\em Phys. Rev.}, D22:2227, 1980.

\bibitem{Mohapatra:1986bd}
R.~N. Mohapatra and J.~W.~F. Valle.
\newblock {Neutrino Mass and Baryon Number Nonconservation in Superstring
  Models}.
\newblock {\em Phys. Rev.}, D34:1642, 1986.

\bibitem{DeRomeri:2017oxa}
Valentina De~Romeri, Enrique Fernandez-Martinez, Julia Gehrlein, Pedro A.~N.
  Machado, and Viviana Niro.
\newblock {Dark Matter and the elusive $Z'$ in a dynamical Inverse Seesaw
  scenario}.
\newblock {\em JHEP}, 10:169, 2017.

\bibitem{Khachatryan:2016vau}
Georges Aad et~al.
\newblock {Measurements of the Higgs boson production and decay rates and
  constraints on its couplings from a} {combined ATLAS and CMS analysis of the
  LHC pp collision data at $ \sqrt{s}=7 $ and 8 TeV}.
\newblock {\em JHEP}, 08:045, 2016.

\bibitem{Ade:2015xua}
P.~A.~R. Ade et~al.
\newblock {Planck 2015 results. XIII. Cosmological parameters}.
\newblock {\em Astron. Astrophys.}, 594:A13, 2016.

\bibitem{Basse:2013zua}
Tobias Basse, Ole~Eggers Bjaelde, Jan Hamann, Steen Hannestad, and Yvonne Y.~Y.
  Wong.
\newblock {Dark energy properties from large future galaxy surveys}.
\newblock {\em JCAP}, 1405:021, 2014.

\bibitem{Amendola:2016saw}
Luca Amendola et~al.
\newblock {Cosmology and Fundamental Physics with the Euclid Satellite}.
\newblock 2016.

\bibitem{Khachatryan:2016zqb}
Vardan Khachatryan et~al.
\newblock {Search for narrow resonances in dilepton mass spectra in
  proton-proton collisions at $\sqrt{s}$ = 13 TeV} {and combination with 8 TeV
  data}.
\newblock {\em Phys. Lett.}, B768:57--80, 2017.

\bibitem{ATLAS:2016cyf}
The~ATLAS collaboration.
\newblock {Search for new high-mass resonances in the dilepton final state
  using proton-proton collisions at $\sqrt{s}$ = 13 TeV with the ATLAS
  detector}.
\newblock 2016.

\bibitem{Akerib:2016vxi}
D.~S. Akerib et~al.
\newblock {Results from a search for dark matter in the complete LUX exposure}.
\newblock {\em Phys. Rev. Lett.}, 118(2):021303, 2017.

\bibitem{Aprile:2015uzo}
E.~Aprile et~al.
\newblock {Physics reach of the XENON1T dark matter experiment}.
\newblock {\em JCAP}, 1604(04):027, 2016.

\bibitem{Akerib:2015cja}
D.~S. Akerib et~al.
\newblock {LUX-ZEPLIN (LZ) Conceptual Design Report}.
\newblock 2015.

\end{thebibliography}



\providecommand{\href}[2]{#2}\begingroup\raggedright\begin{thebibliography}{100}

\bibitem{Tortola:2013voa}
M.~Tortola, \emph{{Status of three-neutrino oscillation parameters}},
  \href{https://doi.org/10.1002/prop.201200106}{\emph{Fortsch. Phys.}
  {\bfseries 61} (2013) 427--440}.

\bibitem{Forero:2014bxa}
D.~V. Forero, M.~Tortola and J.~W.~F. Valle, \emph{{Neutrino oscillations
  refitted}}, \href{https://doi.org/10.1103/PhysRevD.90.093006}{\emph{Phys.
  Rev.} {\bfseries D90} (2014) 093006},
  [\href{https://arxiv.org/abs/1405.7540}{{\ttfamily 1405.7540}}].

\bibitem{Gonzalez-Garcia:2014bfa}
M.~C. Gonzalez-Garcia, M.~Maltoni and T.~Schwetz, \emph{{Updated fit to three
  neutrino mixing: status of leptonic CP violation}},
  \href{https://doi.org/10.1007/JHEP11(2014)052}{\emph{JHEP} {\bfseries 11}
  (2014) 052}, [\href{https://arxiv.org/abs/1409.5439}{{\ttfamily 1409.5439}}].

\bibitem{Gonzalez-Garcia:2015qrr}
M.~C. Gonzalez-Garcia, M.~Maltoni and T.~Schwetz, \emph{{Global Analyses of
  Neutrino Oscillation Experiments}},
  \href{https://doi.org/10.1016/j.nuclphysb.2016.02.033}{\emph{Nucl. Phys.}
  {\bfseries B908} (2016) 199--217},
  [\href{https://arxiv.org/abs/1512.06856}{{\ttfamily 1512.06856}}].

\bibitem{Capozzi:2016rtj}
F.~Capozzi, E.~Lisi, A.~Marrone, D.~Montanino and A.~Palazzo, \emph{{Neutrino
  masses and mixings: Status of known and unknown $3\nu$ parameters}},
  \href{https://doi.org/10.1016/j.nuclphysb.2016.02.016}{\emph{Nucl. Phys.}
  {\bfseries B908} (2016) 218--234},
  [\href{https://arxiv.org/abs/1601.07777}{{\ttfamily 1601.07777}}].

\bibitem{Esteban:2016qun}
I.~Esteban, M.~C. Gonzalez-Garcia, M.~Maltoni, I.~Martinez-Soler and
  T.~Schwetz, \emph{{Updated fit to three neutrino mixing: exploring the
  accelerator-reactor complementarity}},
  \href{https://doi.org/10.1007/JHEP01(2017)087}{\emph{JHEP} {\bfseries 01}
  (2017) 087}, [\href{https://arxiv.org/abs/1611.01514}{{\ttfamily
  1611.01514}}].

\bibitem{Minkowski:1977sc}
P.~Minkowski, \emph{{$\mu \to e\gamma$ at a Rate of One Out of $10^{9}$ Muon
  Decays?}}, \href{https://doi.org/10.1016/0370-2693(77)90435-X}{\emph{Phys.
  Lett.} {\bfseries B67} (1977) 421--428}.

\bibitem{Ramond:1979py}
P.~Ramond, \emph{{The Family Group in Grand Unified Theories}},  in
  \emph{{International Symposium on Fundamentals of Quantum Theory} {and
  Quantum Field Theory Palm Coast, Florida, February 25-March 2, 1979}},
  pp.~265--280, 1979, \href{https://arxiv.org/abs/hep-ph/9809459}{{\ttfamily
  hep-ph/9809459}}.

\bibitem{GellMann:1980vs}
M.~Gell-Mann, P.~Ramond and R.~Slansky, \emph{{Complex Spinors and Unified
  Theories}}, {\emph{Conf. Proc.} {\bfseries C790927} (1979) 315--321},
  [\href{https://arxiv.org/abs/1306.4669}{{\ttfamily 1306.4669}}].

\bibitem{Yanagida:1979as}
T.~Yanagida, \emph{{HORIZONTAL SYMMETRY AND MASSES OF NEUTRINOS}}, {\emph{Conf.
  Proc.} {\bfseries C7902131} (1979) 95--99}.

\bibitem{Mohapatra:1979ia}
R.~N. Mohapatra and G.~Senjanovic, \emph{{Neutrino Mass and Spontaneous Parity
  Violation}}, \href{https://doi.org/10.1103/PhysRevLett.44.912}{\emph{Phys.
  Rev. Lett.} {\bfseries 44} (1980) 912}.

\bibitem{Schechter:1980gr}
J.~Schechter and J.~W.~F. Valle, \emph{{Neutrino Masses in SU(2) x U(1)
  Theories}}, \href{https://doi.org/10.1103/PhysRevD.22.2227}{\emph{Phys. Rev.}
  {\bfseries D22} (1980) 2227}.

\bibitem{Mohapatra:1986bd}
R.~N. Mohapatra and J.~W.~F. Valle, \emph{{Neutrino Mass and Baryon Number
  Nonconservation in Superstring Models}},
  \href{https://doi.org/10.1103/PhysRevD.34.1642}{\emph{Phys. Rev.} {\bfseries
  D34} (1986) 1642}.

\bibitem{Akhmedov:1995vm}
E.~K. Akhmedov, M.~Lindner, E.~Schnapka and J.~W.~F. Valle, \emph{{Dynamical
  left-right symmetry breaking}},
  \href{https://doi.org/10.1103/PhysRevD.53.2752}{\emph{Phys. Rev.} {\bfseries
  D53} (1996) 2752--2780},
  [\href{https://arxiv.org/abs/hep-ph/9509255}{{\ttfamily hep-ph/9509255}}].

\bibitem{Malinsky:2005bi}
M.~Malinsky, J.~C. Romao and J.~W.~F. Valle, \emph{{Novel supersymmetric SO(10)
  seesaw mechanism}},
  \href{https://doi.org/10.1103/PhysRevLett.95.161801}{\emph{Phys. Rev. Lett.}
  {\bfseries 95} (2005) 161801},
  [\href{https://arxiv.org/abs/hep-ph/0506296}{{\ttfamily hep-ph/0506296}}].

\bibitem{Mohapatra:1986aw}
R.~N. Mohapatra, \emph{{Mechanism for Understanding Small Neutrino Mass in
  Superstring Theories}},
  \href{https://doi.org/10.1103/PhysRevLett.56.561}{\emph{Phys. Rev. Lett.}
  {\bfseries 56} (1986) 561--563}.

\bibitem{Roncadelli:1983ty}
M.~Roncadelli and D.~Wyler, \emph{{Naturally Light Dirac Neutrinos in Gauge
  Theories}}, \href{https://doi.org/10.1016/0370-2693(83)90156-9}{\emph{Phys.
  Lett.} {\bfseries B133} (1983) 325--329}.

\bibitem{Roy:1983be}
P.~Roy and O.~U. Shanker, \emph{{Observable Neutrino Dirac Mass and Supergrand
  Unification}}, \href{https://doi.org/10.1103/PhysRevLett.52.713}{\emph{Phys.
  Rev. Lett.} {\bfseries 52} (1984) 713--716}.

\bibitem{Bazzocchi:2010dt}
F.~Bazzocchi, \emph{{Minimal Dynamical Inverse See Saw}},
  \href{https://doi.org/10.1103/PhysRevD.83.093009}{\emph{Phys. Rev.}
  {\bfseries D83} (2011) 093009},
  [\href{https://arxiv.org/abs/1011.6299}{{\ttfamily 1011.6299}}].

\bibitem{Khalil:2010iu}
S.~Khalil, \emph{{TeV-scale gauged B-L symmetry with inverse seesaw
  mechanism}}, \href{https://doi.org/10.1103/PhysRevD.82.077702}{\emph{Phys.
  Rev.} {\bfseries D82} (2010) 077702},
  [\href{https://arxiv.org/abs/1004.0013}{{\ttfamily 1004.0013}}].

\bibitem{Basso:2012ti}
L.~Basso, O.~Fischer and J.~J. van~der Bij, \emph{{Natural Z′ model with an
  inverse seesaw mechanism and leptonic dark matter}},
  \href{https://doi.org/10.1103/PhysRevD.87.035015}{\emph{Phys. Rev.}
  {\bfseries D87} (2013) 035015},
  [\href{https://arxiv.org/abs/1207.3250}{{\ttfamily 1207.3250}}].

\bibitem{Ma:2014qra}
E.~Ma and R.~Srivastava, \emph{{Dirac or inverse seesaw neutrino masses with
  $B-L$ gauge symmetry and $S_3$ flavor symmetry}},
  \href{https://doi.org/10.1016/j.physletb.2014.12.049}{\emph{Phys. Lett.}
  {\bfseries B741} (2015) 217--222},
  [\href{https://arxiv.org/abs/1411.5042}{{\ttfamily 1411.5042}}].

\bibitem{Ma:2015raa}
E.~Ma and R.~Srivastava, \emph{{Dirac or inverse seesaw neutrino masses from
  gauged $B--L$ symmetry}},
  \href{https://doi.org/10.1142/S0217732315300207}{\emph{Mod. Phys. Lett.}
  {\bfseries A30} (2015) 1530020},
  [\href{https://arxiv.org/abs/1504.00111}{{\ttfamily 1504.00111}}].

\bibitem{Escudero:2016tzx}
M.~Escudero, N.~Rius and V.~Sanz, \emph{{Sterile neutrino portal to Dark Matter
  I: The $U(1)_{B-L}$ case}},
  \href{https://doi.org/10.1007/JHEP02(2017)045}{\emph{JHEP} {\bfseries 02}
  (2017) 045}, [\href{https://arxiv.org/abs/1606.01258}{{\ttfamily
  1606.01258}}].

\bibitem{Escudero:2016ksa}
M.~Escudero, N.~Rius and V.~Sanz, \emph{{Sterile Neutrino portal to Dark Matter
  II: Exact Dark symmetry}},
  \href{https://arxiv.org/abs/1607.02373}{{\ttfamily 1607.02373}}.

\bibitem{Abada:2014zra}
A.~Abada, G.~Arcadi and M.~Lucente, \emph{{Dark Matter in the minimal Inverse
  Seesaw mechanism}},
  \href{https://doi.org/10.1088/1475-7516/2014/10/001}{\emph{JCAP} {\bfseries
  1410} (2014) 001}, [\href{https://arxiv.org/abs/1406.6556}{{\ttfamily
  1406.6556}}].

\bibitem{Rojas:2017sih}
N.~Rojas, R.~A. Lineros and F.~Gonzalez-Canales, \emph{{Majoron dark matter
  from a spontaneous inverse seesaw model}},
  \href{https://arxiv.org/abs/1703.03416}{{\ttfamily 1703.03416}}.

\bibitem{Bertone:2004pz}
G.~Bertone, D.~Hooper and J.~Silk, \emph{{Particle dark matter: Evidence,
  candidates and constraints}},
  \href{https://doi.org/10.1016/j.physrep.2004.08.031}{\emph{Phys. Rept.}
  {\bfseries 405} (2005) 279--390},
  [\href{https://arxiv.org/abs/hep-ph/0404175}{{\ttfamily hep-ph/0404175}}].

\bibitem{Bertone:2010zza}
J.~Silk et~al., \emph{{Particle Dark Matter: Observations, Models and
  Searches}}.
\newblock Cambridge Univ. Press, Cambridge, 2010,
  \href{https://doi.org/10.1017/CBO9780511770739}{10.1017/CBO9780511770739}.

\bibitem{Wyler:1982dd}
D.~Wyler and L.~Wolfenstein, \emph{{Massless Neutrinos in Left-Right Symmetric
  Models}}, \href{https://doi.org/10.1016/0550-3213(83)90482-0}{\emph{Nucl.
  Phys.} {\bfseries B218} (1983) 205--214}.

\bibitem{Valle:1982yw}
J.~W.~F. Valle, \emph{{Neutrinoless Double Beta Decay With Quasi Dirac
  Neutrinos}}, \href{https://doi.org/10.1103/PhysRevD.27.1672}{\emph{Phys.
  Rev.} {\bfseries D27} (1983) 1672--1674}.

\bibitem{Valle:1983dk}
J.~W.~F. Valle and M.~Singer, \emph{{Lepton Number Violation With Quasi Dirac
  Neutrinos}}, \href{https://doi.org/10.1103/PhysRevD.28.540}{\emph{Phys. Rev.}
  {\bfseries D28} (1983) 540}.

\bibitem{Weinberg:1979sa}
S.~Weinberg, \emph{{Baryon and Lepton Nonconserving Processes}},
  \href{https://doi.org/10.1103/PhysRevLett.43.1566}{\emph{Phys. Rev. Lett.}
  {\bfseries 43} (1979) 1566--1570}.

\bibitem{Casas:2004gh}
J.~A. Casas, J.~R. Espinosa and I.~Hidalgo, \emph{{Implications for new physics
  from fine-tuning arguments. 1. Application to SUSY and seesaw cases}},
  \href{https://doi.org/10.1088/1126-6708/2004/11/057}{\emph{JHEP} {\bfseries
  11} (2004) 057}, [\href{https://arxiv.org/abs/hep-ph/0410298}{{\ttfamily
  hep-ph/0410298}}].

\bibitem{Vissani:1997ys}
F.~Vissani, \emph{{Do experiments suggest a hierarchy problem?}},
  \href{https://doi.org/10.1103/PhysRevD.57.7027}{\emph{Phys. Rev.} {\bfseries
  D57} (1998) 7027--7030},
  [\href{https://arxiv.org/abs/hep-ph/9709409}{{\ttfamily hep-ph/9709409}}].

\bibitem{Bazzocchi:2009kc}
F.~Bazzocchi, D.~G. Cerdeno, C.~Munoz and J.~W.~F. Valle, \emph{{Calculable
  inverse-seesaw neutrino masses in supersymmetry}},
  \href{https://doi.org/10.1103/PhysRevD.81.051701}{\emph{Phys. Rev.}
  {\bfseries D81} (2010) 051701},
  [\href{https://arxiv.org/abs/0907.1262}{{\ttfamily 0907.1262}}].

\bibitem{Mohapatra:1980qe}
R.~N. Mohapatra and R.~E. Marshak, \emph{{Local B-L Symmetry of Electroweak
  Interactions, Majorana Neutrinos and Neutron Oscillations}},
  \href{https://doi.org/10.1103/PhysRevLett.44.1316}{\emph{Phys. Rev. Lett.}
  {\bfseries 44} (1980) 1316--1319}.

\bibitem{Marshak:1979fm}
R.~E. Marshak and R.~N. Mohapatra, \emph{{Quark - Lepton Symmetry and B-L as
  the U(1) Generator of the Electroweak Symmetry Group}},
  \href{https://doi.org/10.1016/0370-2693(80)90436-0}{\emph{Phys. Lett.}
  {\bfseries B91} (1980) 222--224}.

\bibitem{Pati:1974yy}
J.~C. Pati and A.~Salam, \emph{{Lepton Number as the Fourth Color}},
  \href{https://doi.org/10.1103/PhysRevD.10.275,
  10.1103/PhysRevD.11.703.2}{\emph{Phys. Rev.} {\bfseries D10} (1974)
  275--289}.

\bibitem{Georgi:1974my}
H.~Georgi, \emph{{The State of the Art---Gauge Theories}},
  \href{https://doi.org/10.1063/1.2947450}{\emph{AIP Conf. Proc.} {\bfseries
  23} (1975) 575--582}.

\bibitem{Fritzsch:1974nn}
H.~Fritzsch and P.~Minkowski, \emph{{Unified Interactions of Leptons and
  Hadrons}}, \href{https://doi.org/10.1016/0003-4916(75)90211-0}{\emph{Annals
  Phys.} {\bfseries 93} (1975) 193--266}.

\bibitem{Abada:2014vea}
A.~Abada and M.~Lucente, \emph{{Looking for the minimal inverse seesaw
  realisation}},
  \href{https://doi.org/10.1016/j.nuclphysb.2014.06.003}{\emph{Nucl. Phys.}
  {\bfseries B885} (2014) 651--678},
  [\href{https://arxiv.org/abs/1401.1507}{{\ttfamily 1401.1507}}].

\bibitem{Adhikari:2016bei}
M.~Drewes et~al., \emph{{A White Paper on keV Sterile Neutrino Dark Matter}},
  \href{https://doi.org/10.1088/1475-7516/2017/01/025}{\emph{JCAP} {\bfseries
  1701} (2017) 025}, [\href{https://arxiv.org/abs/1602.04816}{{\ttfamily
  1602.04816}}].

\bibitem{Mohapatra:1980yp}
R.~N. Mohapatra and G.~Senjanovic, \emph{{Neutrino Masses and Mixings in Gauge
  Models with Spontaneous Parity Violation}},
  \href{https://doi.org/10.1103/PhysRevD.23.165}{\emph{Phys. Rev.} {\bfseries
  D23} (1981) 165}.

\bibitem{Lazarides:1980nt}
G.~Lazarides, Q.~Shafi and C.~Wetterich, \emph{{Proton Lifetime and Fermion
  Masses in an SO(10) Model}},
  \href{https://doi.org/10.1016/0550-3213(81)90354-0}{\emph{Nucl. Phys.}
  {\bfseries B181} (1981) 287--300}.

\bibitem{Magg:1980ut}
M.~Magg and C.~Wetterich, \emph{{Neutrino Mass Problem and Gauge Hierarchy}},
  \href{https://doi.org/10.1016/0370-2693(80)90825-4}{\emph{Phys. Lett.}
  {\bfseries 94B} (1980) 61--64}.

\bibitem{Robens:2016xkb}
T.~Robens and T.~Stefaniak, \emph{{LHC Benchmark Scenarios for the Real Higgs
  Singlet Extension of the Standard Model}},
  \href{https://doi.org/10.1140/epjc/s10052-016-4115-8}{\emph{Eur. Phys. J.}
  {\bfseries C76} (2016) 268},
  [\href{https://arxiv.org/abs/1601.07880}{{\ttfamily 1601.07880}}].

\bibitem{Khachatryan:2016vau}
{\scshape ATLAS, CMS} collaboration, G.~Aad et~al., \emph{{Measurements of the
  Higgs boson production and decay rates and constraints on its couplings from
  a} {combined ATLAS and CMS analysis of the LHC pp collision data at $
  \sqrt{s}=7 $ and 8 TeV}},
  \href{https://doi.org/10.1007/JHEP08(2016)045}{\emph{JHEP} {\bfseries 08}
  (2016) 045}, [\href{https://arxiv.org/abs/1606.02266}{{\ttfamily
  1606.02266}}].

\bibitem{Staub:2012pb}
F.~Staub, \emph{{SARAH 3.2: Dirac Gauginos, UFO output, and more}},
  \href{https://doi.org/10.1016/j.cpc.2013.02.019}{\emph{Comput. Phys. Commun.}
  {\bfseries 184} (2013) 1792--1809},
  [\href{https://arxiv.org/abs/1207.0906}{{\ttfamily 1207.0906}}].

\bibitem{Staub:2013tta}
F.~Staub, \emph{{SARAH 4 : A tool for (not only SUSY) model builders}},
  \href{https://doi.org/10.1016/j.cpc.2014.02.018}{\emph{Comput. Phys. Commun.}
  {\bfseries 185} (2014) 1773--1790},
  [\href{https://arxiv.org/abs/1309.7223}{{\ttfamily 1309.7223}}].

\bibitem{Staub:2015kfa}
F.~Staub, \emph{{Exploring new models in all detail with SARAH}},
  \href{https://doi.org/10.1155/2015/840780}{\emph{Adv. High Energy Phys.}
  {\bfseries 2015} (2015) 840780},
  [\href{https://arxiv.org/abs/1503.04200}{{\ttfamily 1503.04200}}].

\bibitem{Vicente:2015zba}
A.~Vicente, \emph{{Computer tools in particle physics}},
  \href{https://arxiv.org/abs/1507.06349}{{\ttfamily 1507.06349}}.

\bibitem{Porod:2003um}
W.~Porod, \emph{{SPheno, a program for calculating supersymmetric spectra, SUSY
  particle decays and SUSY particle} {production at e+ e- colliders}},
  \href{https://doi.org/10.1016/S0010-4655(03)00222-4}{\emph{Comput. Phys.
  Commun.} {\bfseries 153} (2003) 275--315},
  [\href{https://arxiv.org/abs/hep-ph/0301101}{{\ttfamily hep-ph/0301101}}].

\bibitem{Porod:2011nf}
W.~Porod and F.~Staub, \emph{{SPheno 3.1: Extensions including flavour,
  CP-phases and models beyond the MSSM}},
  \href{https://doi.org/10.1016/j.cpc.2012.05.021}{\emph{Comput. Phys. Commun.}
  {\bfseries 183} (2012) 2458--2469},
  [\href{https://arxiv.org/abs/1104.1573}{{\ttfamily 1104.1573}}].

\bibitem{Belyaev:2012qa}
A.~Belyaev, N.~D. Christensen and A.~Pukhov, \emph{{CalcHEP 3.4 for collider
  physics within and beyond the Standard Model}},
  \href{https://doi.org/10.1016/j.cpc.2013.01.014}{\emph{Comput. Phys. Commun.}
  {\bfseries 184} (2013) 1729--1769},
  [\href{https://arxiv.org/abs/1207.6082}{{\ttfamily 1207.6082}}].

\bibitem{Belanger:2014vza}
G.~B{\'e}langer, F.~Boudjema, A.~Pukhov and A.~Semenov, \emph{{micrOMEGAs4.1:
  two dark matter candidates}},
  \href{https://doi.org/10.1016/j.cpc.2015.03.003}{\emph{Comput. Phys. Commun.}
  {\bfseries 192} (2015) 322--329},
  [\href{https://arxiv.org/abs/1407.6129}{{\ttfamily 1407.6129}}].

\bibitem{Alves:2015mua}
A.~Alves, A.~Berlin, S.~Profumo and F.~S. Queiroz, \emph{{Dirac-fermionic dark
  matter in U(1)$_{X}$ models}},
  \href{https://doi.org/10.1007/JHEP10(2015)076}{\emph{JHEP} {\bfseries 10}
  (2015) 076}, [\href{https://arxiv.org/abs/1506.06767}{{\ttfamily
  1506.06767}}].

\bibitem{Lindner:2010rr}
M.~Lindner, A.~Merle and V.~Niro, \emph{{Enhancing Dark Matter Annihilation
  into Neutrinos}},
  \href{https://doi.org/10.1103/PhysRevD.82.123529}{\emph{Phys. Rev.}
  {\bfseries D82} (2010) 123529},
  [\href{https://arxiv.org/abs/1005.3116}{{\ttfamily 1005.3116}}].

\bibitem{Kolb:1990vq}
E.~W. Kolb and M.~S. Turner, \emph{{The Early Universe}}, {\emph{Front. Phys.}
  {\bfseries 69} (1990) 1--547}.

\bibitem{Ade:2015xua}
{\scshape Planck} collaboration, P.~A.~R. Ade et~al., \emph{{Planck 2015
  results. XIII. Cosmological parameters}},
  \href{https://doi.org/10.1051/0004-6361/201525830}{\emph{Astron. Astrophys.}
  {\bfseries 594} (2016) A13},
  [\href{https://arxiv.org/abs/1502.01589}{{\ttfamily 1502.01589}}].

\bibitem{Olive:2016xmw}
{\scshape Particle Data Group} collaboration, C.~Patrignani et~al.,
  \emph{{Review of Particle Physics}},
  \href{https://doi.org/10.1088/1674-1137/40/10/100001}{\emph{Chin. Phys.}
  {\bfseries C40} (2016) 100001}.

\bibitem{Cheung:2012gi}
K.~Cheung, P.-Y. Tseng, Y.-L.~S. Tsai and T.-C. Yuan, \emph{{Global Constraints
  on Effective Dark Matter Interactions: Relic Density, Direct Detection,
  Indirect Detection} {and Collider}},
  \href{https://doi.org/10.1088/1475-7516/2012/05/001}{\emph{JCAP} {\bfseries
  1205} (2012) 001}, [\href{https://arxiv.org/abs/1201.3402}{{\ttfamily
  1201.3402}}].

\bibitem{Akerib:2016vxi}
{\scshape LUX} collaboration, D.~S. Akerib et~al., \emph{{Results from a search
  for dark matter in the complete LUX exposure}},
  \href{https://doi.org/10.1103/PhysRevLett.118.021303}{\emph{Phys. Rev. Lett.}
  {\bfseries 118} (2017) 021303},
  [\href{https://arxiv.org/abs/1608.07648}{{\ttfamily 1608.07648}}].

\bibitem{Aprile:2017iyp}
{\scshape XENON} collaboration, E.~Aprile et~al., \emph{{First Dark Matter
  Search Results from the XENON1T Experiment}},
  \href{https://arxiv.org/abs/1705.06655}{{\ttfamily 1705.06655}}.

\bibitem{Aprile:2015uzo}
{\scshape XENON} collaboration, E.~Aprile et~al., \emph{{Physics reach of the
  XENON1T dark matter experiment}},
  \href{https://doi.org/10.1088/1475-7516/2016/04/027}{\emph{JCAP} {\bfseries
  1604} (2016) 027}, [\href{https://arxiv.org/abs/1512.07501}{{\ttfamily
  1512.07501}}].

\bibitem{Akerib:2015cja}
{\scshape LZ} collaboration, D.~S. Akerib et~al., \emph{{LUX-ZEPLIN (LZ)
  Conceptual Design Report}},
  \href{https://arxiv.org/abs/1509.02910}{{\ttfamily 1509.02910}}.

\bibitem{Ackermann:2015zua}
{\scshape Fermi-LAT} collaboration, M.~Ackermann et~al., \emph{{Searching for
  Dark Matter Annihilation from Milky Way Dwarf Spheroidal Galaxies with Six
  Years} {of Fermi Large Area Telescope Data}},
  \href{https://doi.org/10.1103/PhysRevLett.115.231301}{\emph{Phys. Rev. Lett.}
  {\bfseries 115} (2015) 231301},
  [\href{https://arxiv.org/abs/1503.02641}{{\ttfamily 1503.02641}}].

\bibitem{Abramowski:2011hc}
{\scshape H.E.S.S.} collaboration, A.~Abramowski et~al., \emph{{Search for a
  Dark Matter annihilation signal from the Galactic Center halo with H.E.S.S}},
  \href{https://doi.org/10.1103/PhysRevLett.106.161301}{\emph{Phys. Rev. Lett.}
  {\bfseries 106} (2011) 161301},
  [\href{https://arxiv.org/abs/1103.3266}{{\ttfamily 1103.3266}}].

\bibitem{Wood:2013taa}
M.~Wood, J.~Buckley, S.~Digel, S.~Funk, D.~Nieto and M.~A. Sanchez-Conde,
  \emph{{Prospects for Indirect Detection of Dark Matter with CTA}},  in
  \emph{{Proceedings, Community Summer Study 2013: Snowmass on the Mississippi
  (CSS2013):} {Minneapolis, MN, USA, July 29-August 6, 2013}}, 2013,
  \href{https://arxiv.org/abs/1305.0302}{{\ttfamily 1305.0302}},
  \href{http://www.slac.stanford.edu/econf/C1307292/docs/submittedArxivFiles/1305.0302.pdf}{http://www.slac.stanford.edu/econf/C1307292/docs/submittedArxivFiles/1305.0302.pdf}.

\bibitem{Basse:2013zua}
T.~Basse, O.~E. Bjaelde, J.~Hamann, S.~Hannestad and Y.~Y.~Y. Wong, \emph{{Dark
  energy properties from large future galaxy surveys}},
  \href{https://doi.org/10.1088/1475-7516/2014/05/021}{\emph{JCAP} {\bfseries
  1405} (2014) 021}, [\href{https://arxiv.org/abs/1304.2321}{{\ttfamily
  1304.2321}}].

\bibitem{Amendola:2016saw}
L.~Amendola et~al., \emph{{Cosmology and Fundamental Physics with the Euclid
  Satellite}},  \href{https://arxiv.org/abs/1606.00180}{{\ttfamily
  1606.00180}}.

\bibitem{Khachatryan:2016zqb}
{\scshape CMS} collaboration, V.~Khachatryan et~al., \emph{{Search for narrow
  resonances in dilepton mass spectra in proton-proton collisions at $\sqrt{s}$
  = 13 TeV} {and combination with 8 TeV data}},
  \href{https://doi.org/10.1016/j.physletb.2017.02.010}{\emph{Phys. Lett.}
  {\bfseries B768} (2017) 57--80},
  [\href{https://arxiv.org/abs/1609.05391}{{\ttfamily 1609.05391}}].

\bibitem{ATLAS:2016cyf}
{\scshape ATLAS} collaboration, T.~A. collaboration, \emph{{Search for new
  high-mass resonances in the dilepton final state using proton-proton
  collisions at $\sqrt{s}$ = 13 TeV} {with the ATLAS detector}}, .

\bibitem{Martin:2009iq}
A.~D. Martin, W.~J. Stirling, R.~S. Thorne and G.~Watt, \emph{{Parton
  distributions for the LHC}},
  \href{https://doi.org/10.1140/epjc/s10052-009-1072-5}{\emph{Eur. Phys. J.}
  {\bfseries C63} (2009) 189--285},
  [\href{https://arxiv.org/abs/0901.0002}{{\ttfamily 0901.0002}}].

\bibitem{Dev:2009aw}
P.~S.~B. Dev and R.~N. Mohapatra, \emph{{TeV Scale Inverse Seesaw in SO(10) and
  Leptonic Non-Unitarity Effects}},
  \href{https://doi.org/10.1103/PhysRevD.81.013001}{\emph{Phys. Rev.}
  {\bfseries D81} (2010) 013001},
  [\href{https://arxiv.org/abs/0910.3924}{{\ttfamily 0910.3924}}].

\bibitem{Blanchet:2010kw}
S.~Blanchet, P.~S.~B. Dev and R.~N. Mohapatra, \emph{{Leptogenesis with TeV
  Scale Inverse Seesaw in SO(10)}},
  \href{https://doi.org/10.1103/PhysRevD.82.115025}{\emph{Phys. Rev.}
  {\bfseries D82} (2010) 115025},
  [\href{https://arxiv.org/abs/1010.1471}{{\ttfamily 1010.1471}}].

\bibitem{Abada:2015rta}
A.~Abada, G.~Arcadi, V.~Domcke and M.~Lucente, \emph{{Lepton number violation
  as a key to low-scale leptogenesis}},
  \href{https://doi.org/10.1088/1475-7516/2015/11/041}{\emph{JCAP} {\bfseries
  1511} (2015) 041}, [\href{https://arxiv.org/abs/1507.06215}{{\ttfamily
  1507.06215}}].

\bibitem{Hernandez:2015wna}
P.~Hern{\'a}ndez, M.~Kekic, J.~L{\'o}pez-Pav{\'o}n, J.~Racker and N.~Rius,
  \emph{{Leptogenesis in GeV scale seesaw models}},
  \href{https://doi.org/10.1007/JHEP10(2015)067}{\emph{JHEP} {\bfseries 10}
  (2015) 067}, [\href{https://arxiv.org/abs/1508.03676}{{\ttfamily
  1508.03676}}].

\bibitem{Shrock:1980vy}
R.~E. Shrock, \emph{{New Tests For, and Bounds On, Neutrino Masses and Lepton
  Mixing}}, \href{https://doi.org/10.1016/0370-2693(80)90235-X}{\emph{Phys.
  Lett.} {\bfseries B96} (1980) 159--164}.

\bibitem{Shrock:1980ct}
R.~E. Shrock, \emph{{General Theory of Weak Leptonic and Semileptonic Decays.
  1. Leptonic Pseudoscalar Meson Decays,} {with Associated Tests For, and
  Bounds on, Neutrino Masses and Lepton Mixing}},
  \href{https://doi.org/10.1103/PhysRevD.24.1232}{\emph{Phys. Rev.} {\bfseries
  D24} (1981) 1232}.

\bibitem{Shrock:1981wq}
R.~E. Shrock, \emph{{General Theory of Weak Processes Involving Neutrinos. 2.
  Pure Leptonic Decays}},
  \href{https://doi.org/10.1103/PhysRevD.24.1275}{\emph{Phys. Rev.} {\bfseries
  D24} (1981) 1275}.

\bibitem{Langacker:1988ur}
P.~Langacker and D.~London, \emph{{Mixing Between Ordinary and Exotic
  Fermions}}, \href{https://doi.org/10.1103/PhysRevD.38.886}{\emph{Phys.Rev.}
  {\bfseries D38} (1988) 886}.

\bibitem{Bilenky:1992wv}
S.~M. Bilenky and C.~Giunti, \emph{{Seesaw type mixing and muon-neutrino --->
  tau-neutrino oscillations}},
  \href{https://doi.org/10.1016/0370-2693(93)90760-F}{\emph{Phys.Lett.}
  {\bfseries B300} (1993) 137--140},
  [\href{https://arxiv.org/abs/hep-ph/9211269}{{\ttfamily hep-ph/9211269}}].

\bibitem{Nardi:1994iv}
E.~Nardi, E.~Roulet and D.~Tommasini, \emph{{Limits on neutrino mixing with new
  heavy particles}},
  \href{https://doi.org/10.1016/0370-2693(94)90736-6}{\emph{Phys.Lett.}
  {\bfseries B327} (1994) 319--326},
  [\href{https://arxiv.org/abs/hep-ph/9402224}{{\ttfamily hep-ph/9402224}}].

\bibitem{Tommasini:1995ii}
D.~Tommasini, G.~Barenboim, J.~Bernabeu and C.~Jarlskog, \emph{{Nondecoupling
  of heavy neutrinos and lepton flavor violation}},
  \href{https://doi.org/10.1016/0550-3213(95)00201-3}{\emph{Nucl.Phys.}
  {\bfseries B444} (1995) 451--467},
  [\href{https://arxiv.org/abs/hep-ph/9503228}{{\ttfamily hep-ph/9503228}}].

\bibitem{Antusch:2006vwa}
S.~Antusch, C.~Biggio, E.~Fernandez-Martinez, M.~Gavela and J.~Lopez-Pavon,
  \emph{{Unitarity of the Leptonic Mixing Matrix}},
  \href{https://doi.org/10.1088/1126-6708/2006/10/084}{\emph{JHEP} {\bfseries
  0610} (2006) 084}, [\href{https://arxiv.org/abs/hep-ph/0607020}{{\ttfamily
  hep-ph/0607020}}].

\bibitem{Antusch:2008tz}
S.~Antusch, J.~P. Baumann and E.~Fernandez-Martinez, \emph{{Non-Standard
  Neutrino Interactions with Matter from Physics Beyond the Standard Model}},
  \href{https://doi.org/10.1016/j.nuclphysb.2008.11.018}{\emph{Nucl.Phys.}
  {\bfseries B810} (2009) 369--388},
  [\href{https://arxiv.org/abs/0807.1003}{{\ttfamily 0807.1003}}].

\bibitem{Biggio:2008in}
C.~Biggio, \emph{{The Contribution of fermionic seesaws to the anomalous
  magnetic moment of leptons}},
  \href{https://doi.org/10.1016/j.physletb.2008.09.004}{\emph{Phys. Lett.}
  {\bfseries B668} (2008) 378--384},
  [\href{https://arxiv.org/abs/0806.2558}{{\ttfamily 0806.2558}}].

\bibitem{Forero:2011pc}
D.~V. Forero, S.~Morisi, M.~Tortola and J.~W.~F. Valle, \emph{{Lepton flavor
  violation and non-unitary lepton mixing in low-scale type-I seesaw}},
  \href{https://doi.org/10.1007/JHEP09(2011)142}{\emph{JHEP} {\bfseries 09}
  (2011) 142}, [\href{https://arxiv.org/abs/1107.6009}{{\ttfamily 1107.6009}}].

\bibitem{Abdallah:2011ew}
W.~Abdallah, A.~Awad, S.~Khalil and H.~Okada, \emph{{Muon Anomalous Magnetic
  Moment and mu -> e gamma in B-L Model with Inverse Seesaw}},
  \href{https://doi.org/10.1140/epjc/s10052-012-2108-9}{\emph{Eur. Phys. J.}
  {\bfseries C72} (2012) 2108},
  [\href{https://arxiv.org/abs/1105.1047}{{\ttfamily 1105.1047}}].

\bibitem{Alonso:2012ji}
R.~Alonso, M.~Dhen, M.~Gavela and T.~Hambye, \emph{{Muon conversion to electron
  in nuclei in type-I seesaw models}},
  \href{https://doi.org/10.1007/JHEP01(2013)118}{\emph{JHEP} {\bfseries 1301}
  (2013) 118}, [\href{https://arxiv.org/abs/1209.2679}{{\ttfamily 1209.2679}}].

\bibitem{Abada:2014nwa}
A.~Abada, V.~De~Romeri and A.~M. Teixeira, \emph{{Effect of steriles states on
  lepton magnetic moments and neutrinoless double beta decay}},
  \href{https://doi.org/10.1007/JHEP09(2014)074}{\emph{JHEP} {\bfseries 09}
  (2014) 074}, [\href{https://arxiv.org/abs/1406.6978}{{\ttfamily 1406.6978}}].

\bibitem{Abada:2014cca}
A.~Abada, V.~De~Romeri, S.~Monteil, J.~Orloff and A.~M. Teixeira,
  \emph{{Indirect searches for sterile neutrinos at a high-luminosity
  Z-factory}}, \href{https://doi.org/10.1007/JHEP04(2015)051}{\emph{JHEP}
  {\bfseries 04} (2015) 051},
  [\href{https://arxiv.org/abs/1412.6322}{{\ttfamily 1412.6322}}].

\bibitem{Arganda:2014dta}
E.~Arganda, M.~J. Herrero, X.~Marcano and C.~Weiland, \emph{{Imprints of
  massive inverse seesaw model neutrinos in lepton flavor violating Higgs boson
  decays}}, \href{https://doi.org/10.1103/PhysRevD.91.015001}{\emph{Phys. Rev.}
  {\bfseries D91} (2015) 015001},
  [\href{https://arxiv.org/abs/1405.4300}{{\ttfamily 1405.4300}}].

\bibitem{Abada:2015trh}
A.~Abada and T.~Toma, \emph{{Electric Dipole Moments of Charged Leptons with
  Sterile Fermions}},
  \href{https://doi.org/10.1007/JHEP02(2016)174}{\emph{JHEP} {\bfseries 02}
  (2016) 174}, [\href{https://arxiv.org/abs/1511.03265}{{\ttfamily
  1511.03265}}].

\bibitem{Abada:2016awd}
A.~Abada and T.~Toma, \emph{{Electron electric dipole moment in Inverse Seesaw
  models}}, \href{https://doi.org/10.1007/JHEP08(2016)079}{\emph{JHEP}
  {\bfseries 08} (2016) 079},
  [\href{https://arxiv.org/abs/1605.07643}{{\ttfamily 1605.07643}}].

\bibitem{Abada:2015oba}
A.~Abada, V.~De~Romeri and A.~M. Teixeira, \emph{{Impact of sterile neutrinos
  on nuclear-assisted cLFV processes}},
  \href{https://doi.org/10.1007/JHEP02(2016)083}{\emph{JHEP} {\bfseries 02}
  (2016) 083}, [\href{https://arxiv.org/abs/1510.06657}{{\ttfamily
  1510.06657}}].

\bibitem{Abada:2015zea}
A.~Abada, D.~Be{\v c}irevi{\'c}, M.~Lucente and O.~Sumensari, \emph{{Lepton
  flavor violating decays of vector quarkonia and of the $Z$ boson}},
  \href{https://doi.org/10.1103/PhysRevD.91.113013}{\emph{Phys. Rev.}
  {\bfseries D91} (2015) 113013},
  [\href{https://arxiv.org/abs/1503.04159}{{\ttfamily 1503.04159}}].

\bibitem{Fernandez-Martinez:2015hxa}
E.~Fernandez-Martinez, J.~Hernandez-Garcia, J.~Lopez-Pavon and M.~Lucente,
  \emph{{Loop level constraints on Seesaw neutrino mixing}},
  \href{https://doi.org/10.1007/JHEP10(2015)130}{\emph{JHEP} {\bfseries 10}
  (2015) 130}, [\href{https://arxiv.org/abs/1508.03051}{{\ttfamily
  1508.03051}}].

\bibitem{DeRomeri:2016gum}
V.~De~Romeri, M.~J. Herrero, X.~Marcano and F.~Scarcella, \emph{{Lepton flavor
  violating Z decays: A promising window to low scale seesaw neutrinos}},
  \href{https://doi.org/10.1103/PhysRevD.95.075028}{\emph{Phys. Rev.}
  {\bfseries D95} (2017) 075028},
  [\href{https://arxiv.org/abs/1607.05257}{{\ttfamily 1607.05257}}].

\bibitem{Abada:2016vzu}
A.~Abada, V.~De~Romeri, J.~Orloff and A.~M. Teixeira, \emph{{In-flight cLFV
  conversion: $e-\mu$, $e-\tau$ and $\mu-\tau$ in minimal extensions of the
  Standard Model} {with sterile fermions}},
  \href{https://doi.org/10.1140/epjc/s10052-017-4864-z}{\emph{Eur. Phys. J.}
  {\bfseries C77} (2017) 304},
  [\href{https://arxiv.org/abs/1612.05548}{{\ttfamily 1612.05548}}].

\bibitem{Antusch:2014woa}
S.~Antusch and O.~Fischer, \emph{{Non-unitarity of the leptonic mixing matrix:
  Present bounds and future sensitivities}},
  \href{https://doi.org/10.1007/JHEP10(2014)094}{\emph{JHEP} {\bfseries 1410}
  (2014) 94}, [\href{https://arxiv.org/abs/1407.6607}{{\ttfamily 1407.6607}}].

\bibitem{Fernandez-Martinez:2016lgt}
E.~Fernandez-Martinez, J.~Hernandez-Garcia and J.~Lopez-Pavon, \emph{{Global
  constraints on heavy neutrino mixing}},
  \href{https://doi.org/10.1007/JHEP08(2016)033}{\emph{JHEP} {\bfseries 08}
  (2016) 033}, [\href{https://arxiv.org/abs/1605.08774}{{\ttfamily
  1605.08774}}].

\end{thebibliography}\endgroup

\end{document}